\documentclass[11pt]{article}
\usepackage{amsmath}
\usepackage{amssymb}
\usepackage{bbm}
\usepackage{epsfig}
\allowdisplaybreaks[4]
\def\pa{\partial}
\def\besub{\begin{subequations}}
\def\eesub{\end{subequations}}
\def\be{\begin{equation}}
\def\ee{\end{equation}}
\def\bea{\begin{eqnarray}}
\def\eea{\end{eqnarray}}
\def\nn{\nonumber\\}

\def\STr{{\rm STr }}
\def\Tr{{\rm Tr }}

\newcommand{\eqn}[1]{Eq.~(\ref{#1})}
\newcommand{\eqns}[2]{Eqs.~(\ref{#1}),(\ref{#2})}
\newcommand{\eqnss}[3]{Eqs.~(\ref{#1}),(\ref{#2}),(\ref{#3})}
\newcommand{\reference}[1]{Ref.~\cite{#1}}

\begin{document}
\title{Bounds on scalar masses in two Higgs doublet models}
\author{P.M. Ferreira$^{1,2}$~\footnote{ferreira@cii.fc.ul.pt},
D.R.T. Jones$^{3}$~\footnote{drtj@sune.amtp.liv.ac.uk} \\
$^1$ Instituto Superior de Engenharia de Lisboa, \\ Rua Conselheiro
Em\'{\i}dio
Navarro, 1, 1959-007 Lisboa, Portugal \\
$^2$ Centro de F\'{\i}sica Te\'orica e Computacional, Faculdade de Ci\^encias,\\
Universidade de Lisboa, Av. Prof. Gama Pinto, 2, 1649-003 Lisboa,
Portugal \\
$^3$ Department of Mathematical Sciences, University of Liverpool,
\\
Liverpool, L69 3BX, UK}
\date{\today}
\maketitle \noindent {\bf Abstract.} A thorough analysis of
stability and perturbativity bounds is performed in several versions
of the two-Higgs doublet model, for both CP-conserving and
spontaneously broken CP minima. LEP results further aid in
establishing very strict constraints on the mass of the lighter
Higgs particle.

\vspace{1cm}

\section{Introduction}

Despite the great successes of the Standard Model (SM) of particle
physics, it leaves many unanswered questions, such as the origin of
matter-antimatter asymmetry; although the SM does contain  a
CP-violating parameter in the CKM matrix, and violates baryon number, it
is generally accepted that  it does not lead to baryogenesis sufficient
to  explain the observed asymmetry. One of the simplest extensions of
the SM, which tries to solve this problem, is the two-Higgs doublet
model (2HDM)~\cite{lee}, wherein a second Higgs doublet is added to the
theory. The spectrum of scalar particles becomes richer and, for some
realisations of the model, spontaneous breaking of the CP symmetry is
possible. The 2HDM presents some challenges, though: except in
supersymmetric models, the quartic interactions between the scalar
doublets are not theoretically constrained, and increase substantially
the number of free parameters.  As a consequence the predictive power of
the model is reduced. Any tool available to constrain the parameter
space is thus of great interest. In this paper we will take a closer
look at the requirements of stability and perturbativity of the model.
Namely, we will analyse their impact on the several possible
incarnations of the model;  including versions involving the imposition
of  two types of global symmetries, which eliminate several unknown
parameters. For various reasons, it may be of interest to  break those
symmetries softly, by the introduction of quadratic coefficients in the
potential. The possible different vacua of the model - minima which
spontaneously break CP or preserve it - require a separate stability
and perturbativity analysis, which we will perform. This paper is
organised as follows: in Section~\ref{sec:2hdm} we will briefly review
the basic notions about the 2HDM scalar potential and the requirements
of stability and perturbativity for a range of renormalisation scales.
This will lead to the computation of the one-loop $\beta$-functions of
the model, (previously given in \reference{Haber:1993an}). In
sections~\ref{sec:z2} to~\ref{sec:10} we will apply the stability and
perturbativity bounds to the several realisations of the 2HDM: models
with a discrete $Z_2$ or global $U(1)$ symmetries and their softly
broken counterparts; within these, we will consider the possible cases
of minima with spontaneously broken CP, or unbroken CP; and finally we
will also consider the most general CP-conserving 2HDM potential. In all
cases, we will endeavour to obtain bounds on the masses of the scalar
particles, and use the latest experimental results on Higgs searches
from LEP~\cite{lep} to further constrain the potential's parameter
space. Details of the $\beta$-function calculation are given in
Appendix~\ref{sec:ap}, following a simple and pedagogical approach which
may be of interest for readers unfamiliar with it;  and in
Appendix~\ref{sec:aph}\ we make some remarks about  the renormalisation
group invariance of basis-invariant conditions on the  couplings.

\section{The 2HDM potential}
\label{sec:2hdm}

The 2HDM potential~\cite{lee} involves two Higgs doublets with
hypercharge $Y\,=\,1$, $\Phi_1$ and $\Phi_2$, and is invariant under
the gauge symmetries of the standard model, $SU(3)_C \times SU(2)_W
\times U(1)_Y$. The most general potential one can build with these
two doublets, following the conventions of~\cite{hab}, is given by
\begin{align}
V\;= &\;
m^2_{11}\,\Phi_1^\dagger\Phi_1\,+\,m^2_{22}\,\Phi_2^\dagger\Phi_2\,-\,\left(
m^2_{12}\,\Phi_1^\dagger\Phi_2\,+\,\mbox{h.c.}\right) \,+ \nonumber \\
 &\; \frac{\lambda_1}{2}\,\left(\Phi_1^\dagger\Phi_1\right)^2\,+\,
\frac{\lambda_2}{2}\,\left(\Phi_2^\dagger\Phi_2\right)^2\,+\,
\lambda_3\,\left(\Phi_1^\dagger\Phi_1\right)\,\left(\Phi_2^\dagger\Phi_2\right)\,+
\,
\lambda_4\,\left(\Phi_1^\dagger\Phi_2\right)\,\left(\Phi_2^\dagger\Phi_1\right)\,+
\nonumber \\
 &\; \left\{\frac{\lambda_5}{2}\,\left(\Phi_1^\dagger\Phi_2\right)^2\,+\,
 \left[\lambda_6\,\left(\Phi_1^\dagger\Phi_1\right)\,+\,
 \lambda_7\,\left(\Phi_2^\dagger\Phi_2\right) \right]\,
 \left(\Phi_1^\dagger\Phi_2\right) \,+\,\mbox{h.c.}\right\} \;\;\; ,
 \label{eq:pot}
\end{align}
where the couplings
$\{m^2_{12}\,,\,\lambda_5\,,\,\lambda_6\,,\,\lambda_7\}$ are in
general complex. In all, this potential has at most 14 real
parameters. If one defines the CP transformation of the scalar
fields as $\Phi_1\,\rightarrow\,\Phi_1^*$,
$\Phi_2\,\rightarrow\,\Phi_2^*$, and requires that the potential
above preserves this symmetry, all the parameters become necessarily
real. The number of free real parameters is thus reduced to 10,
or in fact 9, with an appropriate choice of basis for the scalar
doublets. Namely, a given linear combination of $\Phi_1$ and
$\Phi_2$ will always diagonalise the quadratic terms in the fields
in Eq.~\eqref{eq:pot}. For all that follows, we will consider that
the potential does not break CP explicitly, and thus all parameters
are taken as real.

In general both $\Phi_1$ and $\Phi_2$ could have distinct Yukawa
couplings to up-type quarks,  down-type quarks and leptons. However,
these generic Yukawa couplings would induce flavour-changing neutral
currents (FCNC) in the theory. These have to be kept in check,
either by imposing severe bounds on the size of the model's
parameters or, more elegantly, by imposing symmetries upon it.
Namely, a discrete $Z_2$ symmetry~\cite{sim} or a global
$U(1)$~\cite{PQ} will prevent any FCNC from arising. This can be
accomplished in several ways, but the choice we will make in this
paper is to have only $\Phi_1$ coupling to fermions. The results
for the $\beta$-functions of the model are easily
generalised to other situations, by means of the techniques detailed
in Appendix~\ref{sec:ap}. In what follows we will, however, retain
only the top quark Yukawa coupling, as the remaining ones will be
too small to have any meaningful effect on the analysis we will
perform. With this assumption we present in Appendix~\ref{sec:ap}
the one loop $\beta$-functions for the theory defined by
\eqn{eq:pot}. Expressions for the $\beta$-functions for
general models may be found in the literature~\cite{betas},
and the explicit expressions including the
$\lambda_{6,7}$ contributions were given in \reference{Haber:1993an}.
Even with FCNC-preventing symmetries imposed upon it, the 2HDM
potential has a great number of free parameters - 7 or more of them
- a fact which severely curtails its predictive power. Any tools
which help in limiting this vast parameter space are thus welcome.
One way to limit the values of the quartic couplings
in~\eqn{eq:pot} is by observing that general values for the
$\lambda_i$ do not guarantee that the potential is bounded from
below (BFB). In fact, lest one requires that the quartic terms
in~\eqn{eq:pot} do not tend to minus infinity for any direction in
field space, we will have no guarantees that the potential can have
a stable minimum. For potentials where
$\lambda_6\,=\,\lambda_7\,=\,0$, Ivanov~\cite{iva} has proven that
the 2HDM potential is bounded from below if and only if the
following conditions are obeyed:
\begin{align}
\lambda_1 & >\, 0 & \lambda_3 & >\,-\,\sqrt{\lambda_1 \lambda_2} \nonumber \\
\lambda_2 & >\, 0 & \lambda_3\,+\,\lambda_4\,-\,|\lambda_5| &
>\,-\,\sqrt{\lambda_1 \lambda_2}  \;\;\; . \label{eq:unb1}
\end{align}
These conditions have been widely used in the literature and assumed
to be only {\em necessary} ones, but they are also in fact {\em
sufficient}. The work of~\cite{iva} gives, in principle, all
necessary and sufficient conditions to have the potential
Eq.~\eqref{eq:pot} bounded from below even in the case
$\lambda_6\,\neq\,0$, $\lambda_7\,\neq\,0$, but the relations one
could derive in this situation are extremely complicated, and not at
all clear. See also~\cite{kli}. In~\cite{eu} {\em necessary}
conditions involving $\lambda_6$ and $\lambda_7$ were derived, and
we will use them in the following work:
\begin{equation}
2\,|\lambda_6\,+\,\lambda_7|\;<\;\frac{\lambda_1\,+\,\lambda_2}{2}\,+\,
\lambda_3\,+\,\lambda_4\,+\,\lambda_5\;\;\; . \label{eq:unb2}
\end{equation}

As explained above, the conditions Eqs.~\eqref{eq:unb1}
and~\eqref{eq:unb2} ensure the stability of the tree scalar potential.
To be sure of a viable vacuum, however, one must
take into account the effect of radiative corrections, and the related fact that
the $\lambda_i$ depend on the renormalisation scale $\mu$.

Let us first review this important issue in the context of a theory
with a single scalar field, the real scalar $\phi^4$-model, with \be
V_{cl}=\frac{m^2}{2} \phi^2+\frac{\lambda}{4!}\phi^4 \ee so that \be
V = V_{cl} +\frac{V_{cl}''(\phi)^2}{64\pi^2}
\ln\left(\frac{V_{cl}''(\phi)}{\mu^2}\right)+ \cdots \ee Let us
suppose for simplicity that $m^2 > 0$. For what values of $\phi$ can
we reliably calculate $V$?  Suppose we have chosen a RG scale $\mu
\sim m$ and  that $\lambda$ is small on that scale so that
perturbation theory in $\lambda$  is believable. Then evidently  we
can calculate $V$ as  $\phi\to 0$ by  simply retaining $\mu \sim m$
since the one loop correction  is obviously small. Thus the origin
remains a minimum, as was the case for the tree potential.
But what about $\phi > > m$? The one-loop
correction  now becomes large, because of the logarithm, so that one
must improve on this perturbation expansion.  RG improvement amounts, in
fact, to  exploiting the freedom to choose the renormalisation scale to
take $\mu^2 \sim V_{cl}''(\phi)$, or $\mu \sim \phi$ for large $\phi$.
Then to a good approximation  at large  $\phi$ we will have
\be
V = \frac{\lambda (\phi)}{4!}\phi^4
\ee
and this will be perturbatively believable as long as $\lambda (\phi)$ is small.
Now in this simple model $\lambda$ becomes large at large
scales, approaching a Landau pole, and so perturbation breaks down
eventually in spite of our RG improvement. Thus we cannot say what form the
potential takes at sufficiently large $\phi$.

In a more complicated theory there are two main issues to take into
consideration.  Firstly,  if the potential depends on more than one
scalar field, it is not immediately obvious in which directions in
field space we will be able to describe the large-field potential,
since we have only one scale at our disposal\footnote{For an attempt
to generalise the RG discussion to incorporate more than one $\mu$
see \reference{Einhorn:1983fc}.}. Secondly, the behaviour of the
$\lambda_i (\phi)$  for large $\phi$ may be quite different from
that in the simple $\phi^4$-model. In particular, in the 2HDM the
large size of the top quark Yukawa coupling, and the sign of its
contribution to the $\beta$-function of $\lambda_1$ (see
Eq.~\eqref{eq:fullbetas} in Appendix~\ref{sec:ap}) drives down the
value of that quartic coupling as one goes up in renormalization
scale. If the starting point of $\lambda_1$ is sufficiently small,
$\lambda_1$ may become negative at a given high value of $\mu$,
which would mean that any minimum present for low renormalization
scales would in fact be unstable - the potential would either  be
unbounded from below or develop a much deeper minimum  at large
$\phi_1$\footnote{In fact, in the type of  theory we consider here,
the latter is generally the case because the positive contribution
of the gauge coupling contributions to  their $\beta$-functions
causes  $\lambda_i$ to recover to positive values at yet higher
scales.}.

The approach we shall take is to simply assume that the stability
conditions \eqns{eq:unb1}{eq:unb2} must hold at all renormalisation
scales $\mu$ up to the (putative) gauge unification scale
$M_U\,=\,10^{15}$ GeV. This will clearly be sufficient to produce a
potential bounded from below. Requiring the {\em stability} of the
scalar potential at all scales will thus, typically, impose {\em
lower} bounds on the values of its quartic couplings.

Another way of limiting the values of the $\lambda_i$ is by
requiring that they remain small enough for perturbative
believability at high scales. Hence, if the initial values of
$\lambda_i$ are too large, their $\beta$-functions will be positive
and their renormalization scale evolution will drive them to ever
higher values. Requiring that the $\lambda_i$ remain small at all
scales will thus impose {\em upper} bounds on their values. How
small should ``small" be? Here we enter a somewhat arbitrary region,
but requiring that all $\lambda_i$ remain less than 10 at all
renormalization scales seems a reasonable requirement.

We will therefore impose both stability and perturbativity bounds on
the quartic parameters of the 2HDM at all scales between the weak
scale $M_Z$ and $M_U$. Such analyses have been made
before, in many works: these ideas were applied to the
SM~\cite{spsm}, SUSY models~\cite{spsu} and also to a simple
2HDM~\cite{sp2h}. In this work we are interested in studying the
differences that the application of these bounds will have on the
several possible two-Higgs doublet models, and on the several
possible vacua therein possible.

\section{Model with $Z_2$ symmetry}
\label{sec:z2}

One of the symmetries that rids the potential Eq.~\eqref{eq:pot} of
FCNC was first proposed by Glashow, Weinberg and Paschos~\cite{sim},
and consists of a simple $Z_2$ transformation in the fields:
$\Phi_1\,\rightarrow\,-\,\Phi_1$, $\Phi_2\,\rightarrow\,\Phi_2$. By
carefully choosing similar transformations for the fermionic fields
it is possible to eliminate the existence of FCNCs by having, for
instance, only $\Phi_1$ couple to the fermions. This symmetry
simplifies Eq.~\eqref{eq:pot}, namely setting to zero several of the
couplings: $m_{12} \,=\,\lambda_6\,=\,\lambda_7\,=\,0$. Then the BFB
conditions of Eq.~\eqref{eq:unb1} are, in this case, necessary and
sufficient.

Our procedure was as follows: we generated many thousands of
combinations of quartic parameters of the 2HDM. The couplings were
generated with magnitudes between $10^{-3}$ and $10$, allowing
different couplings to have different orders of magnitude and to be
negative if allowed. Spontaneous symmetry breaking occurs when the
doublets acquire vacuum expectation values such that
\begin{equation}
<\Phi_1>\,=\,\begin{pmatrix} 0 \\ v_1 \end{pmatrix}\;\;\; ,\;\;\;
<\Phi_2>\,=\,\begin{pmatrix} 0 \\ v_2 \end{pmatrix}\;\;\;.
\label{eq:vevs}
\end{equation}
We thus generated values for the vevs $\{v_1\,,\,v_2\}$ such that
$v_1^2\,+\,v_2^2\,=\,v^2$, with $v\,=\, 246/\sqrt{2}$ GeV. With the
vevs and all quartic couplings, it is simple to use the stationarity
conditions of the model and determine the quadratic parameters
$m_{11}^2$ and $m_{22}^2$. At this point, we have a full set of
parameters for the potential. By analysing the model's squared
scalar mass matrices (expressions for which may be found, for
instance, in~\cite{eu}), we can ensure that each combination of
parameter values under consideration is indeed a Normal minimum of
the 2HDM.

\begin{figure}[ht]
\epsfysize=10cm \centerline{\epsfbox{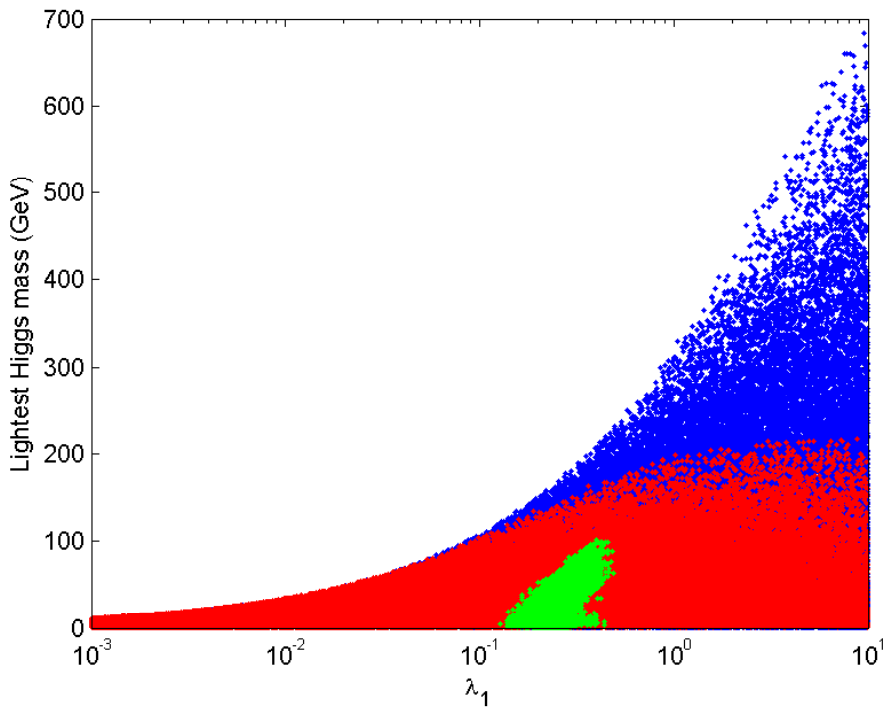}} \caption{Results of
scan of the 2HDM potential with $Z_2$ symmetry.} \label{fig:z2_g}
\end{figure}

\begin{figure}[ht]
\epsfysize=10cm \centerline{\epsfbox{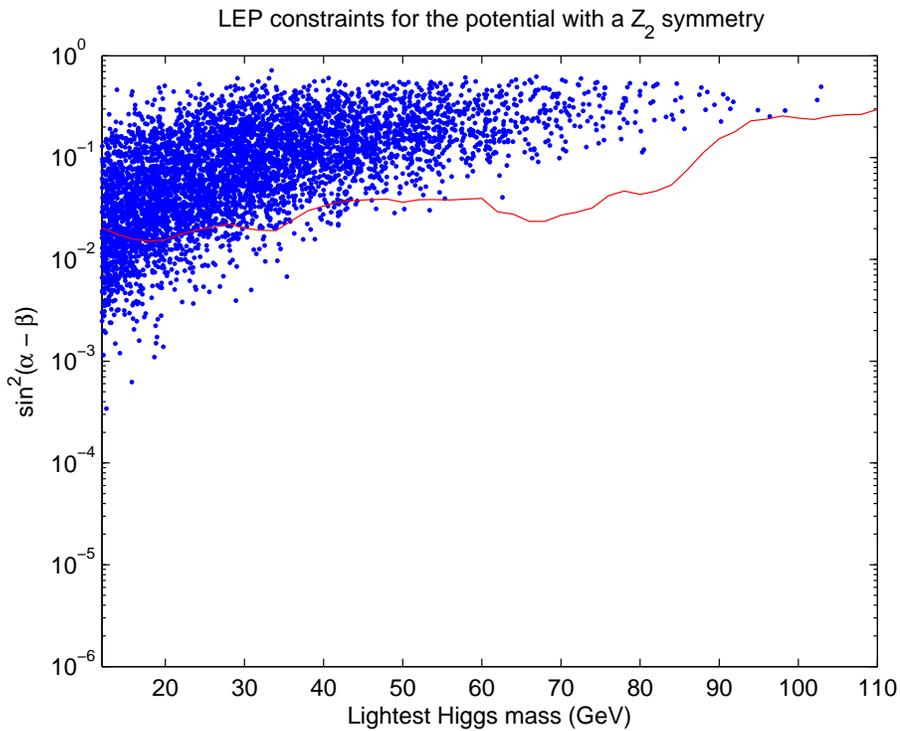}}
\caption{$\sin^2(\alpha-\beta)$ versus the mass of the lightest
Higgs boson, for a potential with a $Z_2$ symmetry.}
\label{fig:xsiz2}
\end{figure}

We then analysed the RG evolution of the quartic couplings for each
``point" of parameter space and checked whether they obeyed the
stability and triviality bounds described above, between $M_Z$
and $M_U$. In Fig.~\ref{fig:z2_g} we see the result of this
procedure. In this
plot we show the lightest CP-even Higgs mass versus the value of the
$\lambda_1$ coupling. The colours are interpreted as:
\begin{itemize}
\item The red (medium) points represent those combinations
of 2HDM parameters
for which the stability conditions of Eq.~\eqref{eq:unb1} were violated
somewhere between $M_Z$ and $M_U$.
\item The blue (dark) points represent all of the parameter combinations
which passed the stability conditions of Eq.~\eqref{eq:unb1}, but for which
no triviality conditions were set.
\item Finally, the green (lightest) points are a subset of the blue ones -
those for which the triviality conditions are obeyed at all
scales between $M_Z$ and $M_U$~\footnote{Notice that there are many blue points
``between" the red ones, even if in the plot they are ``covered" by the red points.
This means that not all points with Higgs mass below $\sim$ 200 GeV
are excluded on the basis of the stability conditions, and that the green region is
indeed a subset of the blue one. This is not a contradiction - this parameter space includes
seven different parameters, and this plot is only varying two. We can therefore have a
``rejected" point and an ``accepted" one occupying the same place in the plot.}.
\end{itemize}
As we see, the combination of stability and triviality conditions
narrows the ``allowed" range of $\lambda_1$ immensely - the only
values which ``survive" are in the interval
$0.24\,<\,\lambda_1\,<\,0.91$. The remaining couplings are likewise
constrained in similar intervals, of identical order of magnitude.
This also limits high values for the Higgs scalar masses. In fact,
if one analyses the full spectrum of scalar particles, one finds
that after the stability and triviality requirements the masses are
bounded by (roughly):
\begin{align}
m_h & <\, 102 \;\;\mbox{GeV} \nonumber \\
121\,<\,m_H & <\, 199 \;\;\mbox{GeV} \nonumber \\
m_A & <\, 163 \;\;\mbox{GeV} \nonumber \\
m_{H^\pm} & <\, 160 \;\;\mbox{GeV} \;\;\; . \label{eq:bounds}
\end{align}
These results, specially those pertaining to the charged Higgs mass,
are in agreement with previous works (see the last reference of
~\cite{sp2h}). Unless explicitly stated, no lower bounds were found
for these masses, the exception being the heaviest neutral scalar
$H$. These bounds do not preclude very low Higgs masses, then. In
fact, current experimental data does not forbid light neutral 2HDM
scalars. The best bounds on Higgs masses arise from the latest LEP
results~\cite{lep} and the analysis of associated production of a Z
boson with the lightest 2HDM CP-even scalar, through the triple
vertex $Z\,Z\,h$. In the 2HDM, the coupling associated with this
vertex is equal to its SM value, multiplied by $\sin(\alpha-\beta)$,
where $\tan\beta = v_2/v_1$ and $\alpha$ is the usual mixing angle
for the CP-even scalar mass matrix. If this coupling is small -
meaning, if $\sin(\alpha-\beta)$ is small - then the mass of the
scalar particle can be also small and still have escaped detection
at LEP. This was explored recently for both SUSY models and one
version of the 2HDM (see, for instance,~\cite{xsi}). In fact, the
cross section for $e^+\,e^-\,\rightarrow\,Z\,h$ production in the
2HDM is related to that of the SM by
\begin{equation}
\sigma^{2HDM}(e^+ e^- \rightarrow Z h)\,=\,\sin^2(\alpha-\beta)\,
\sigma^{SM}(e^+ e^- \rightarrow Z h) \;\;\; . \label{eq:sig}
\end{equation}
This relation is valid for any type of 2HDM model with a
CP-conserving vacuum. The LEP results impose severe constraints on
the size of the ratio $\sigma^{2HDM}/\sigma^{SM}$ which, considering
the previous equation, translate as constraints on
$\sin^2(\alpha-\beta)$. In Fig.~\ref{fig:xsiz2} we plot the value of
$\sin^2(\alpha-\beta)$ against the mass of the lightest Higgs boson,
for the subset of parameter space which survived the stability and
triviality bounds (the green (light) points from
Fig.~\ref{fig:z2_g}). The
red (continuous) line drawn in this plot corresponds to the experimental bound
coming from the LEP searches~\cite{lep}. Only the points below
this line are allowed. We therefore see that the
majority of the points which survived the triviality and
stability analysis are already excluded on experimental grounds.
Although it seems possible to generate high masses for the lightest
Higgs particle, this plot clearly indicates that lower masses are
preferred. Indeed, a rough upper bound of $\sim$ 55 GeV can be
established from these data.

Nevertheless, caution must be urged. However large our sampling of
the parameter space, it does not cover all regions of it. Also,
these results are sensitive to the input top quark mass, which still has some
uncertainty, according to the most recent Tevatron
results~\cite{teva}: the CDF and D0 combined value is
$M_t\,=\,173.1\,\pm\,1.3$ GeV. The physical mass corresponds
to the pole of the propagator, and its relation to the
Yukawa coupling $h_t$ and the vev $v_1$ is given, up to one loop, by
\begin{equation}
M_t^{pole}\;=\;h_t v_1\,\left[
1\,+\,\left(4\,-\,3\,\ln\left(\frac{h_t^2
v_1^2}{\mu^2}\right)\,\frac{\alpha_S}{3\pi}\right)\right]\;\;\; ,
\end{equation}
where we are only taking the most significant corrections, those
from QCD. $\mu$ is the renormalization scale considered, and all
quantities in the formula above are evaluated at that scale.
The results presented thus far (and elsewhere in this paper) assume
a top pole mass of 173 GeV. We verified what changes occur if we
varied the top pole mass by 2 GeV in either direction (a conservative
variation). The bounds shown in Eq.~\eqref{eq:bounds}
that change by variation of $M_t^{pole}$ are shown in
Table~\ref{tab:bo}.
\begin{table}[h!]
\begin{center}
\renewcommand{\arraystretch}{1.2}
\begin{tabular}{c|ccc}
\hline \hline
 Pole mass (GeV) & $M_t^{pole}\,=\,171$ & $M_t^{pole}\,=\,173$   &
 $M_t^{pole}\,=\,175$  \\
\hline
             & $m_h \,<\, 100$ & $m_h \,<\, 102$  & $m_h \,<\, 102$  \\
 Mass bounds & $m_H \,>\, 119$ & $m_H \,>\, 121$  & $m_H \,>\, 129$  \\
 (GeV)       & $m_A \,<\, 165$ & $m_A \,<\, 163$
 & $m_A \,<\, 162$ \\
\hline \hline
\end{tabular}
\end{center}
\label{parameters} \caption{Bounds on scalar masses in function of
the value of the top quark pole mass.} \label{tab:bo}
\end{table}
The lower bound on the heaviest CP-even scalar is the one that changes the
most. In fact, that lower bound correspond to small values of $v_2$, for
which one of the masses, $h$ or $H$, is essentially proportional to $\lambda_1$.
The lower bounds for $m_H$ presented in the table above correspond to the lower
allowed values for $\lambda_1$, which obviously change when the top pole mass
is varied. The uncertainty on the top pole mass is thus relevant, and needs
to be factored in evaluating whatever bounds we will present in this
work.

Still, the results shown in Fig.~\ref{fig:xsiz2} clearly indicate
that the 2HDM with a $Z_2$ symmetry is already severely constrained
by the simultaneous requirements of stability, triviality and
compliance with existing experimental results. The pole mass
dependence is of the order of $\sim$ 5 GeV around the central values
at most, and will not drastically change those conclusions.

\subsection{The case $v_2 = 0$}
\label{sec:v20}

For the 2HDM with an unbroken $Z_2$ symmetry, the minimisation conditions
admit a different type of solution than the one we have been considering:
to wit, a vacuum where one of the fields $\Phi$ has a vanishing expectation
value. These models were first proposed in~\cite{dema} and have been studied
before, in many different contexts. For instance, in ref.~\cite{abjp}, one such
model was used to show that it
was possible to have neutrino mixing even without massive neutrinos. In~\cite{ma}
the model was used to explain low neutrino masses as a loop effect. In general
these models are the basis of the so-called ``inert Higgs" theories~\cite{inert},
which have excellent scalar candidates for dark matter~\cite{dark}.

Since in our models only $\Phi_1$ couples to the fermions, we should
therefore study the case where $v_2\,=\,0$, and verify what changes
occur in the bounds we have deduced~\footnote{We thank the referee for bringing
this point to our attention.}. The expressions for the squared scalar masses
in these models are extremely simple, namely
\begin{align}
m^2_{h_1} &=\; 2\,\lambda_1\,v^2 & m^2_{h_2} &=\;
m^2_{22}\,+\,(\lambda_3 + \lambda_4)\,v^2  \nonumber \\
m^2_{H^\pm} &=\; m^2_{22}\,+\,\lambda_3 \,v^2 & M^2_A &=\; m^2_{h_2} \;\;\; ,
\end{align}
where $m^2_{h_1}$ and $m^2_{h_2}$ are the two CP-even scalar masses, the lightest
of which will be $h$, the heaviest $H$ (depending on the parameters, though, we cannot
{\em a priori} guarantee which of $h_1$ and $h_2$ is the lightest state). In the model of
the previous section, the minimisation conditions ensured a strong bond between the
values of the squared parameters, $m_{11}^2$ and $m^2_{22}$, the scale $v^2$ and the
values of the quartic parameters $\lambda_i$. With $v_2\,=\,0$, though, despite the fact that
$m_{11}^2$ is fixed such that $m_{11}^2\,=\,-\,\lambda_1\,v^2$, the parameter $m^2_{22}$
is unconstrained and as such can be as large as one wishes. Thus, we expect the upper
bounds on most of the masses written above to be much larger than before. In fact, once the
stability and triviality analysis is concluded, we obtain
\begin{align}
m_h & <\, 235 \;\;\mbox{GeV} \nonumber \\
m_H & >\, 120 \;\;\mbox{GeV}  \;\;\; .
\label{eq:bv20}
\end{align}
The only upper bound that remains is that on $m_h$, certainly due to
the $h_1$ state, which is directly tied to the severely constrained $\lambda_1$ coupling.
Because the CP-even $2\times 2$ mass matrix is diagonal, the mixing angle $\alpha$
has only two possible values: $0$ and $\pi/2$. As such (and since in this model the
angle $\beta$ is equal to zero), the LEP results have no impact on the model:
\begin{itemize}
\item If $\alpha\,=\,0$, the coupling of $h$ to the $Z$ boson vanishes. This occurs
for a large range of masses, from very low Higgs masses to high ones. In any case,
the LEP data do not provide any constraints, since this lightest scalar does not couple
to the $Z$ and as such could not have been observed at LEP.
\item If $\alpha\,=\,\pi/2$, the coupling of $h$ to $Z$ is identical to that of the SM.
However, this case only occurs for values of the Higgs mass larger than 116 GeV. The LEP
constraints, as can be observed from fig.~\ref{fig:xsiz2}, are only valid for Higss masses
inferior to about 110 GeV. As such, the case $\alpha\,=\,\pi/2$ is also not constrained by
the LEP data.
\end{itemize}

\section{Model with softly broken $Z_2$ symmetry}
\label{sec:8}

If one adds to the $Z_2$ potential a term of the form
$m^2_{12}\,\Phi_1^\dagger\Phi_2\,+\,\mbox{h.c.}$ the discrete
symmetry is softly broken. However, no FCNCs arise from this soft
breaking. The main reason why doing this should be of interest is
quite simple - with this soft breaking term the potential can now
have two types of interesting minima: (a) ``normal" ones, which
preserve CP, and for which the doublet's vacuum expectation values,
as before, have the form of Eq.~\eqref{eq:vevs}; and (b), minima for
which CP is spontaneously broken, the doublets developing vevs of
the form
\begin{equation}
<\Phi_1>\,=\,\begin{pmatrix} 0 \\ v_1 \end{pmatrix}\;\;\; ,\;\;\;
<\Phi_2>\,=\,\begin{pmatrix} 0 \\ v_2  + i v_3\end{pmatrix}\;\;\;.
\label{eq:vevs_cp}
\end{equation}
In order to have minima of phenomenological interest, the CP vevs
must obey $v_1^2\,+\,v_2^2\,+\,v_3^2\,=\,v^2$ GeV$^2$. It has
recently been proved~\cite{iva} that these different minima cannot
co-exist. The parameters of the potential either allow the normal
minimum or a CP breaking one~\footnote{Charge breaking vacua might
also occur but, as shown in~\cite{eu}, whenever a normal minimum
exists in the 2HDM, the global minimum of the potential is normal
and thus safe against charge breaking~\label{f:cb}. Likewise, if a
CP minimum exists, any charge breaking stationary point will
necessarily be a saddle point.}. Thus, the parameter sets for each
of these cases have to be different, and generated separately. The
different vevs, however, do not directly affect the RGE running of
the quartic couplings. Notice also that the existence of another
free parameter ($m^2_{12}$) in the potential will certainly change
the scalar masses.

\subsection{Normal minima}
\label{sec:8n}

For this case, the parameters of the potential were chosen such that
the global minimum of the potential preserves CP. The range of
allowed values for the $\lambda_i$ was the same as before, and the soft
breaking parameter $m_{12}^2$ is chosen so that all three
mass squared parameters are of the order $v^2$, and so that $|m_{12}^2| < 10v^2$.
Once again, thousands of different parameter sets were
generated, and then RGE
analysed between the weak and unification scales. Again we found
that only a narrow window of values of $\lambda_1$ survived the
imposition of stability and triviality bounds. The results
differ from those of the unbroken $Z_2$ model, the bounds found for
the masses being given by
\begin{align}
m_h & <\, 187 \;\;\mbox{GeV} \nonumber \\
121\,<\,m_H & <\, \;\;\mbox{(no upper limit)} \;\;\mbox{GeV}\;\;\; .
\end{align}
As explained earlier, the quadratic parameters $m_{11}^2$ and
$m_{22}^2$ are obtained from the stationarity conditions of the
potential, once all other parameters have been generated. For the
unbroken $Z_2$ potential, those included all the quartic couplings
(limited by the triviality and stability requirements) and the
vevs (limited by the requirement of their squared sum be equal to
$(246/2)^2$ GeV$^2$). Now, however, we have the extra quadratic
parameter in the potential, which is not restricted and can make the
masses larger than they can be in the unbroken $Z_2$ model. This
justifies the lessening of the bounds we discover.

\begin{figure}[ht]
\epsfysize=10cm \centerline{\epsfbox{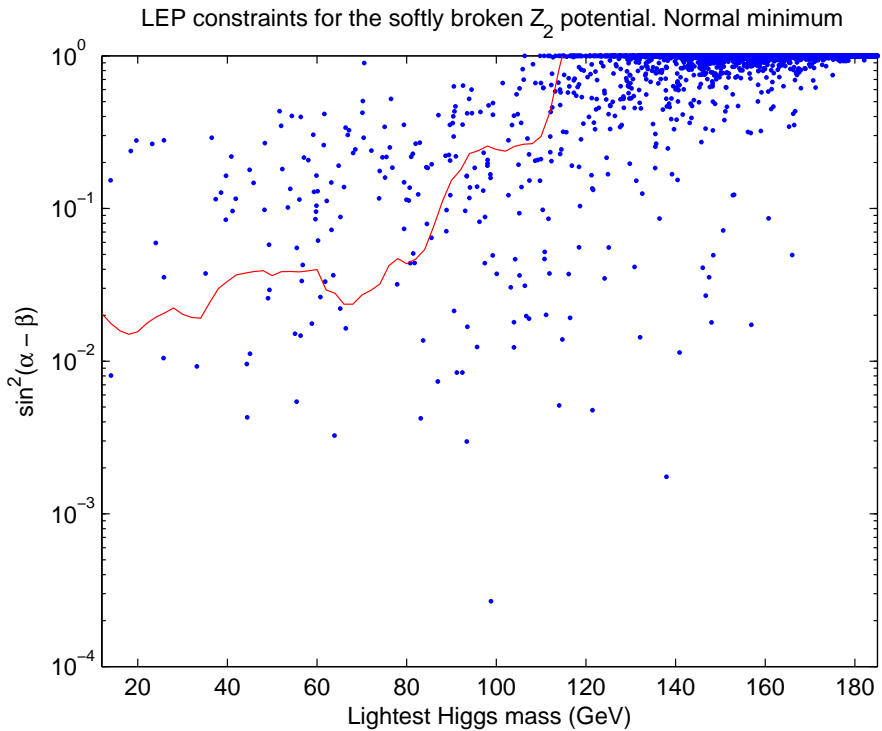}}
\caption{$\sin^2(\alpha-\beta)$ versus the mass of the lightest
Higgs boson, for the normal minimum of a potential with a softly
broken $Z_2$ symmetry.} \label{fig:xsi8n}
\end{figure}

In Fig.~\eqref{fig:xsi8n} we again
plot the effect of the LEP bounds; once again the allowed region is
below (to the right of) the line. As we see, the soft-breaking term
allows us to easily evade the experimental constraints.

\subsection{CP breaking minima}
\label{sec:8cp}

As was shown in Ref.~\cite{eu}, the quartic parameters of the 2HDM
potential need to obey a specific condition so that there is a
CP-breaking minimum\footnote{Namely, the quartic parameters must be
such that the matrix $B_{CP}$ defined in~\cite{eu} be positive
definite, and so that $\lambda_4\,<\,\lambda_5$.}. Thus, if the
quartic parameters are necessarily different from those which
generate Normal minima, one expects different results stemming from
the stability and triviality bounds.

Notice, now, that there is no distinction between CP-even and CP-odd
scalars, since CP is spontaneously broken - the complex vevs shown
in Eq.~\eqref{eq:vevs_cp} will cause a mixing between all neutral
components of the doublets. We rename the neutral scalars as $h_1$,
$h_2$ and $h_3$, in decreasing order of masses. After repeating the
stability and triviality RG analysis, we found the following
bounds for these scalar masses:
\begin{align}
131\,<\,m_{h_1} & <\, 203 \;\;\mbox{GeV} \nonumber \\
6\,<\,m_{h_2} & <\, 160 \;\;\mbox{GeV} \nonumber \\
m_{h_3} & <\, 84 \;\;\mbox{GeV} \nonumber \\
m_{H^\pm} & <\, 158\;\; \mbox{GeV} \;\;\; .
\end{align}
Notice the very low upper bound on the lightest neutral scalar, and
indeed in all of the scalars. Clearly the requirement of a
CP-breaking vacuum ``chooses", from within the parameter space which
survives the stability and triviality bounds, scalars with lower
masses, which is quite different from what we saw occurring for this
same model, for Normal minima. One might speculate that this
phenomenon is somehow related to the Georgi-Pais theorem~\cite{Georgi:1974au}
according to which spontaneous breaking of CP via {\it radiative
corrections} is always
accompanied by scalars which are massless in the tree approximation
(of course here we are considering tree-level CP breaking).
\footnote{We thank the referee for a comment on this point.}

\begin{figure}[ht]
\epsfysize=10cm \centerline{\epsfbox{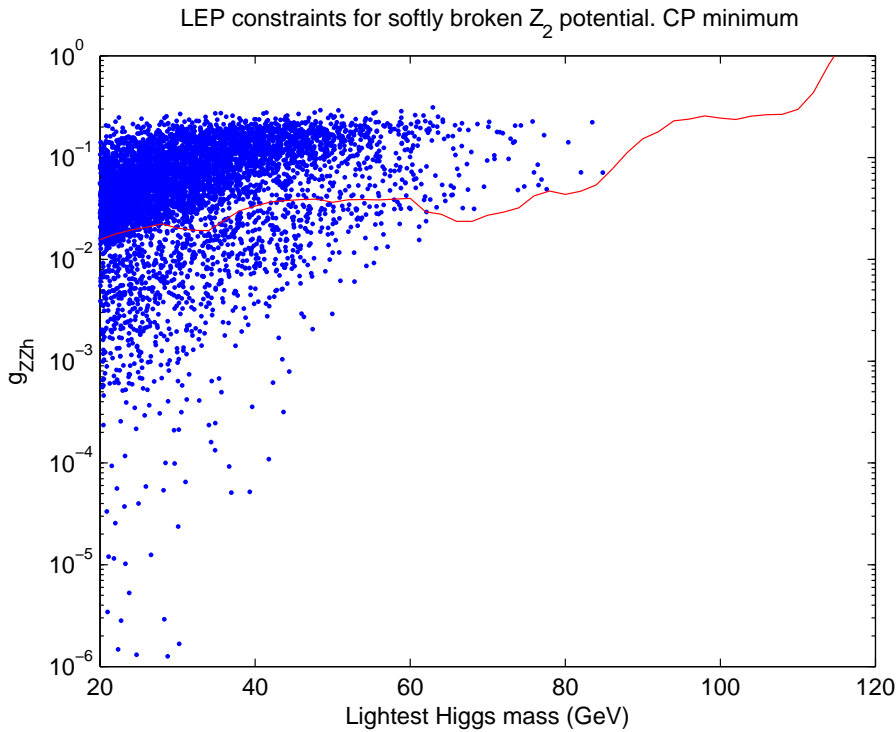}}
\caption{$g_{ZZh}^2$ versus the mass of the lightest Higgs boson,
for the CP breaking minimum of a potential with a softly broken
$Z_2$ symmetry.} \label{fig:xsi8cp}
\end{figure}

The LEP results also have a substantial impact on the parameter
space for these minima. But to apply them, we must first compute the
coupling between the $Z$ boson and the lightest Higgs scalar for a
CP-breaking minimum. As was explained earlier, the complex vevs of
Eq.~\eqref{eq:vevs_cp} cause a mixing between CP-even and CP-odd
scalars, so that the masses of the neutral scalars $h_1$, $h_2$ and
$h_3$ are the eigenvalues of a $4\times 4$ matrix (the fourth
eigenvalue is zero, corresponding to the $Z$ would-be Goldstone mode
$G^0$). Consequently there is now no single angle $\alpha$ which
characterises the diagonalisation of this matrix, and thus the
quantity $\sin(\alpha-\beta)$, which was the ratio between the
$Z\,Z\,h$ coupling in the 2HDM and the SM, is no longer defined for
CP-breaking minima. For the  determination of the $Z\,Z\,h$
coupling, the relevant term in the Lagrangian stems from the kinetic
terms for the doublets, namely
\begin{equation}
\sum_{i=1}^2\,\left(D_\mu\Phi_i\right)^\dagger
\left(D^\mu\Phi_i\right)\;\rightarrow\; \frac{1}{8}
\,g^2\,\sec^2\theta_W\,Z_\mu\,Z^\mu\,\sum_{i=1}^2\,|\Phi_i|^2\;\;\;
. \label{eq:zzh}
\end{equation}
Let us define the neutral component fields of the doublets as
\begin{equation}
\Phi_1^0\,=\,R_1\,+\,i I_1\;\;\; , \;\;\; \Phi_2^0\,=\,R_2\,+\,i I_2
\;\;\; .
\end{equation}
The relationship between these fields and the mass-eigenstates
$h_1$, $h_2$, $h_3$ and $G^0$ is given by a $4\times 4$ unitary
matrix $A_{ij}$, such that
\begin{equation}
\begin{pmatrix} R_1 \\ R_2 \\ I_1 \\ I_2 \end{pmatrix}\;=\;A\,
\begin{pmatrix} h_1 \\ h_2 \\ h_3 \\ G^0 \end{pmatrix} \;\;\;.
\end{equation}
Then, the terms in Eq.~\eqref{eq:zzh} which are directly proportional to
the lightest Higgs, $h_3$, will be given by
\begin{equation}
\frac{1}{4} \,g^2\,\sec^2\theta_W\,Z_\mu\,Z^\mu\,h_3\,\left(A_{13}
v_1\,+\,A_{23} v_2\,+\,A_{43} v_3\right)
\end{equation}
so that the ratio of the $Z\,Z\,h$ coupling in a CP minimum and that
of the SM is given by
\begin{equation}
g_{ZZh}\;=\;\frac{1}{v}\,\left(A_{13} v_1\,+\,A_{23} v_2\,+\,A_{43}
v_3\right)
\end{equation}
with $v\,=\,246/\sqrt{2}$ GeV. Therefore, the generalisation of
Eq.~\eqref{eq:sig} is thus
\begin{equation}
\sigma^{2HDM}_{CP\,\,minimum}(e^+ e^- \rightarrow Z
h)\,=\,g_{ZZh}^2\, \sigma^{SM}(e^+ e^- \rightarrow Z h) \;\;\; .
\label{eq:sigcp}
\end{equation}
It is now a simple task to calculate the mass matrix for the neutral
scalars, diagonalise it and obtain the matrix $A$ and thus compute
the value of $g_{ZZh}$ for the points of parameter space which
survived the stability and triviality bounds, comparing them
with the LEP results. As we see from Fig.~\eqref{fig:xsi8cp}, the
LEP results exclude a significant portion of the parameter space.
(As before the allowed region is below the line). In
fact, according to this plot, the highest value allowed for the
lightest Higgs mass would be around 65 GeV. Comparing this plot to
figs.~\ref{fig:xsiz2} and~\ref{fig:xsi8n}, we see that the bounds we
are imposing have very different consequences, depending on the
model, or type of minimum, considered.

\section{Model with a softly broken $U(1)$ symmetry}
\label{sec:u1}

Another symmetry to eliminate FCNC in the original
potential Eq.~\eqref{eq:pot} is a simple global $U(1)$ transformation in
the fields (accompanied by suitable fermion transformations) of the
form $\Phi_1\,\rightarrow\,\Phi_1$,
$\Phi_2\,\rightarrow\,e^{i\,\alpha}\,\Phi_2$. This symmetry simplifies
Eq.~\eqref{eq:pot}, again setting to zero several of its
couplings, $m_{12}
\,=\,\lambda_5\,=\,\lambda_6\,=\,\lambda_7\,=\,0$. This symmetry is
however {\em too} strong, in that it produces a zero mass axion~\footnote{Unless
one considers a vacuum with $v_2\,=\,0$, as in section~\ref{sec:v20}, with
analogous consequences.}. To prevent that from happening one usually softly
breaks this global $U(1)$ by re-introducing the $m_{12}^2$ term in the potential.
Except for the fact that the remaining $\lambda$ couplings are unrelated to one
another, the resulting Higgs potential is similar to the MSSM
one. The only types of minima this potential possesses are normal
ones; CP breaking is impossible here (see also footnote~\ref{f:cb}).

Once again we generated thousands of parameter sets corresponding to
normal minima and used the model's $\beta$-functions to verify
whether the stability and triviality conditions were satisfied all
the way up to $M_U$. In terms of that analysis, the only difference
with the softly-broken $Z_2$ model studied in section~\ref{sec:8n}
is the fact that for this model we must have $\lambda_5\,=\,0$, even
after the $U(1)$ symmetry has been softly broken. As before, only a
very narrow range of vales of $\lambda_1$ survived the stability and
triviality requirements. The bounds found for the scalar masses
are now:
\begin{align}
m_h & <\, 187 \;\;\mbox{GeV} \nonumber \\
121\,<\,m_H & <\, \;\;\mbox{(no upper limit)} \;\;\mbox{GeV}
\;\;\; .
\end{align}
The LEP constraints do not
\begin{figure}[ht]
\epsfysize=10cm \centerline{\epsfbox{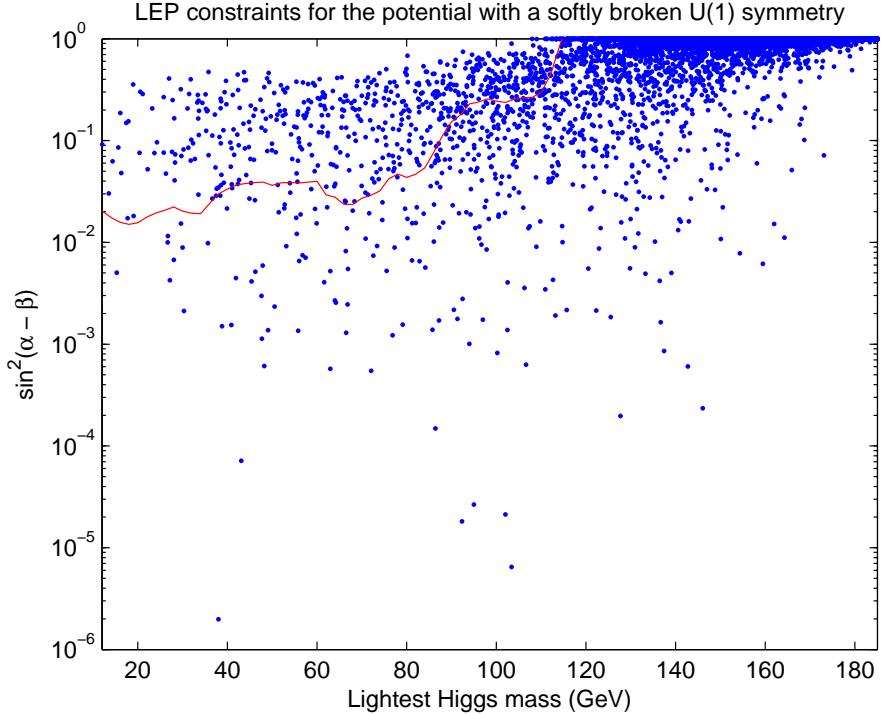}}
\caption{$\sin^2(\alpha-\beta)$ versus the mass of the lightest
Higgs boson, for the CP breaking minimum of a potential with a
softly broken $U(1)$ symmetry.} \label{fig:xsiu1}
\end{figure}
affect the available parameter space as much as in the model with
unbroken $Z_2$ symmetry, or for CP minima in the softly broken $Z_2$
model; the results are shown in Fig.~\eqref{fig:xsiu1}. The results
are quite similar to those obtained in section~\ref{sec:8n} for
normal minima in the softly-broken $Z_2$ model, no doubt due to the
presence of in both cases of the soft-breaking parameter $m_{12}^2$.

\section{The full CP conserving potential}
\label{sec:10}

What about the full potential of Eq.~\eqref{eq:pot}? What do the
triviality and stability bounds tell us about it? To perform
this analysis we need the $\beta$-functions for this model, in terms
of the ``new" couplings $\lambda_6$ and $\lambda_7$. These
are to be found in Appendix~\ref{sec:ap}. As we
have mentioned, in this case FCNC will, in general, occur.

Due to the presence of the new quartic couplings, the stability
analysis needs to take into account Eq.~\eqref{eq:unb2}, which
involves $\lambda_6$ and $\lambda_7$. And once more, as with the
softly broken $Z_2$ model, this potential can have minima with
spontaneous CP breaking and Normal minima, though the same set of
parameters cannot produce two such minima in coexistence. We thus
obtain the following bounds for each type of minima: for Normal
minima, the only bounds found were
\begin{align}
m_h & <\, 187 \;\;\mbox{GeV} \nonumber \\
121\,<\,m_H & <\, \;\;\mbox{(no upper limit)} \;\;\mbox{GeV}
\;\;\; ,
\end{align}
and for CP minima,
\begin{align}
131\,<\,m_{h_1} & <\, 205 \;\;\mbox{GeV} \nonumber \\
6\,<\,m_{h_2} & <\, 166 \;\;\mbox{GeV} \nonumber \\
m_{h_3} & <\, 84 \;\;\mbox{GeV} \nonumber \\
m_{H^\pm} & <\, 158\;\; \mbox{GeV} \;\;\; .
\end{align}
As we can see, there are no great differences with the results
obtained for the Normal minima of the softly broken $Z_2$ potential,
or for the $U(1)$ model. The new parameters do not affect the bounds
on the masses, nor do they change the qualitative difference between
the bounds obtained for each type of minima: that the requirements
of stability and triviality tend to ``pick" lower scalar masses
for CP minima than they do for Normal minima. Also, notice that the case
$v_2\,=\,0$, discussed in section~\ref{sec:v20}, is only possible for the $Z_2$ 
symmetric potential (in fact solutions with $v_2\,=\,0$ become possible as long as 
$m_{12}^2\,=\,\lambda_6\,=\,0$, but without the full $Z_2$ symmetry these 
conditions would not be preserved by renormalisation).
It corresponds to a completely different type of vacuum (one which preserves the
$Z_2$ symmetry) for which the bounds we found for that case (eq.~\eqref{eq:bv20}) are unaffected.

\begin{figure}[ht]
\epsfysize=10cm \centerline{\epsfbox{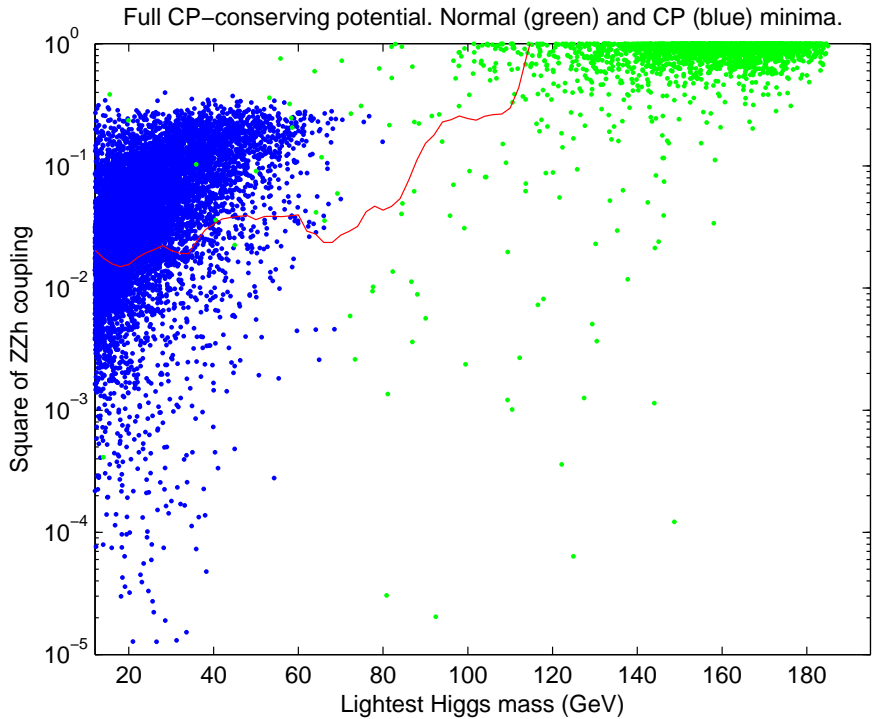}} \caption{Squared $ZZh$
coupling versus the mass of the lightest Higgs boson, for the full
CP conserving potential. The green (light) points concern the Normal minima,
the blue (dark) ones the minima which spontaneously break CP.}
\label{fig:10}
\end{figure}

The LEP constraints are again similar to those already shown, and we present them,
for both minima, in Fig.~\ref{fig:10}.
Once again, much of the available parameter space is excluded by the
LEP data for the case of the CP minima. The lightest Higgs scalar,
in that case, would have a rough upper bound of $\sim$ 80 GeV.

\section{Conclusions}

We have performed a thorough analysis of the impact that the demands
of stability and triviality have on the scalar masses of the
2HDM. We considered several possible incarnations of this model -
models with $Z_2$ or $U(1)$ symmetries, with those symmetries softly
broken or simply without them, and the different neutral vacua
allowed in those theories. At the same time, we studied the impact
of the LEP results on production of a light Higgs scalar on the
parameter space of the model. Our results may be summarised as
follows:
\begin{itemize}
\item The 2HDM potential with a $Z_2$ symmetry is very strongly
constrained. Combining both theoretical and experimental
constraints, the mass of the lightest neutral scalar should be
less than about 55 GeV.
\item The LEP restrictions are easily avoided in Normal minima of models
with softly broken $Z_2$ or $U(1)$ symmetries, or in the full
CP-conserving potential. In those models, the lightest CP-even
neutral scalar is bound to be smaller than about 190 GeV. The
heaviest CP-even neutral scalar is bound to be larger than about 120
GeV.
\item The minima with spontaneous CP breaking which may occur in
these models are heavily constrained. For these minima, the
stability and triviality bounds affect the parameter space of
these models in different ways than what occurs for the Normal
minima. Those bounds constrain very tightly {\em all} scalar masses,
and the LEP results have an extremely strong impact on the surviving
parameter space, eliminating most of it.
\end{itemize}

In order to perform this analysis, we needed the $\beta$-functions
for the parameters $\lambda_6$ and $\lambda_7$, which are
given in Appendix~\ref{sec:ap} and were given previously in~\cite{Haber:1993an}.
These $\beta$-functions allowed us to verify the
validity of necessary conditions involving $\lambda_6$ and
$\lambda_7$ between the weak and renormalization scales. Their
usefulness, however, is not restricted to the studies presented
here. In~\cite{cp}, for instance, basis-invariant conditions which
ensure greater symmetry of the 2HDM potential were obtained. Several
of those conditions involve the parameters $\lambda_6$ and
$\lambda_7$. To verify if these conditions are valid at all
renormalisation scales, it will be necessary to employ the
$\beta$-functions. We perform this analysis in
Appendix~\ref{sec:aph}.

The parameters $\lambda_6$ and $\lambda_7$  are often omitted
because in their presence there  is no global symmetry which we can
use to automatically prevent flavour changing neutral currents. We
have avoided this issue by assuming  that quarks couple to $\Phi_1$
only, and anyway retaining  only the top quark Yukawa coupling. Of
course the absence of a symmetry to  enforce this means that to
pursue the question of FCNCs we  would need to consider the effect
of radiative corrections.  In any event, we found the effect of
$\lambda_{6,7}$ on our  analysis to be limited - the analysis of the
stability and triviality bounds for the full CP-conserving
potential did not produce significantly different results from those
obtained for the softly broken $Z_2$ or $U(1)$ models. That in
itself, however, is an interesting result: if for whatever reason
one wishes to work with the full CP-conserving 2HDM model, one can
do so with the certainty that the $\lambda_6$ and $\lambda_7$
parameters will not spoil the restrictive bounds obtained in simpler
models.

Finally, the bounds we obtained here considered that the stability and
triviality conditions held at all scales between $M_Z$ and $M_U$.
More to the point, this procedure has the underlying assumption that the
two-Higgs doublet model constitutes the whole of physics up to the gauge
unification scale. That philosophy, however, can be readily inverted. A
possible scenario is that within the next few years several scalar
particles are discovered at the LHC, and that their properties conform
to the 2HDM. However, suppose their masses completely violate all bounds
presented here. This would, of course,  suggest the existence of more
new physics beyond the 2HDM  but below the gauge unification scale, to
justify the breaking of the stability and triviality bounds. The
simplest example of such new physics would be the existence of a  heavy
fourth family of fermions, the presence of which would significantly
change the form of the $\beta$-functions of the model,  by virtue of the
(necessarily large) associated Yukawa couplings.

\vspace{0.25cm} {\bf Acknowledgements:} This work was supported in
part by the Portuguese {\em Funda\c{c}\~{a}o para a Ci\^{e}ncia e a
Tecnologia} (FCT) under contract PTDC/FIS/70156/2006. DRTJ also
thanks the CERN Theory Division for hospitality and financial support.

\appendix

\section{The one loop $\beta$-functions}
\label{sec:ap}

In this appendix we describe the calculation of the one loop
$\beta$-functions for the general two Higgs scalar model  defined by
the potential given in \eqn{eq:pot}. The calculation  is
straightforward by normal diagrammatic methods, and has
already been presented in \reference{Haber:1993an}. Here we describe an
alternative algebraic calculation based on the renormalisation group
equation satisfied by the effective potential, which we hope may be
of some pedagogic interest, as well as providing a check
on the previous calculation; in which we indeed thereby
identify one fairly obvious typo. This procedure uses the RG invariance of
the one loop effective potential, where all the $\beta$-functions
may be computed from $STr M^4$, where $M^2$ is the mass matrix of
the fields, including arbitrary vevs for the scalars. The basic
procedure is explained in section 6 of reference~\cite{fjt}.

Up to one loop, the effective potential $V_{\rm eff}(\phi)$ for any
theory is given by:
\be
V_{\rm eff}(\phi) = V(\phi) + V_1 (\phi) + \cdots
\ee
where $V(\phi)$ is the tree potential given in our case
by~\eqn{eq:pot}, and
\be
V_1 (\phi) = \frac{\kappa}{4} \, \STr M^4 \ln
\frac{M^2}{\mu^2}\;\;\;,
\ee
where the mass matrix $M^2$ includes
contributions to all scalar, fermion  and vector boson mass matrices
with arbitrary background values of all scalar fields $\phi$,
$\kappa = (16\pi^2)^{-1}$ and $\STr$ is the usual spin-weighted
trace.

In the Landau gauge,  $V_{\rm eff}$ obeys the following RG equation:
\be
\left[\mu\frac{\pa}{\pa\mu}+\sum_i\beta_i\frac{\pa}{\pa\lambda_i}
-(\phi\gamma\frac{\pa}{\pa\phi}+ {\rm c.c.})\right]V_{\rm eff}=0
\ee
where the $\lambda_i$ include all mass parameters and coupling
constants, and $\gamma$ is the  matrix of anomalous dimensions of
the scalar fields.

It follows that
\be {\cal D}^{(1)}V=-\mu\frac{\pa}{\pa\mu}V_1 = \frac{\kappa}{2} \,
\STr M^4 \;\;\; ,
\label{eq:rgone}
\ee
where
\be
{\cal D}^{(n)}=\sum_i\beta^{(n)}_i\frac{\pa}{\pa\lambda_i}
-\left(\phi\gamma^{(n)}\frac{\pa}{\pa\phi} + {\rm c.c.}\right).
\ee
By comparing coefficients of the various $\phi^4$ terms on the two
sides of \eqn{eq:rgone}\ we can, if we know $\gamma$, determine all
the one-loop $\beta$ functions. From now on we write $\beta_i^{(1)}=
\kappa\beta_i$ and $\gamma^{(1)}= \kappa\gamma$ to avoid writing
factors of $\kappa$.

Let us first consider the simplified case when we set gauge and
Yukawa couplings to zero. Then we can write $M^2$ as follows:
\be
M^2 = \left(\begin{array}{cc}
A & B \\
C & D \end{array}
\right)
\ee
where
\be
A = \left(\begin{array}{cc} \frac{\pa^2
V}{\pa\phi_i\pa\phi^{\dagger j}}
& \frac{\pa^2 V}{\pa\phi_i\pa\xi^{\dagger j}}\\
\frac{\pa^2 V}{\pa\xi_i\pa\phi^{\dagger j}}
& \frac{\pa^2 V}{\pa\xi_i\pa\xi^{\dagger j}} \end{array}
\right)
\ee

\be
B = \left(\begin{array}{cc}
\frac{\pa^2 V}{\pa\phi_i\pa\phi_j} & \frac{\pa^2 V}{\pa\phi_i\pa\xi_j}\\
\frac{\pa^2 V}{\pa\xi_i\pa\phi_j} & \frac{\pa^2 V}{\pa\xi_i\pa\xi_j} \end{array}
\right)
\ee

\be
C = \left(\begin{array}{cc}
\frac{\pa^2 V}{\pa\phi^{\dagger i}\pa\phi^{\dagger j}} &
\frac{\pa^2 V}{\pa\phi^{\dagger i}\pa\xi^{\dagger j}}\\
\frac{\pa^2 V}{\pa\xi^{\dagger i}\pa\phi^{\dagger j}} &
\frac{\pa^2 V}{\pa\xi^{\dagger i}\pa\xi^{\dagger j}} \end{array}
\right)
\ee

\be
D = \left(\begin{array}{cc}
\frac{\pa^2 V}{\pa\phi^{\dagger i}\pa\phi_j}
& \frac{\pa^2 V}{\pa\phi^{\dagger i}\pa\xi_j}\\
\frac{\pa^2 V}{\pa\xi^{\dagger i}\pa\phi_j} & \frac{\pa^2
V}{\pa\xi^{\dagger i}\pa\xi_j} \end{array} \right)
\ee
and $i,j = 1,2$ are $SU(2)$ indices, and to control the profusion of
indices we have put $\Phi_1 \equiv \phi$ and $\Phi_2 \equiv \xi$.

It is then straightforward to write down the matrices $A,B,C$.
Thus, for example,
\bea
(A_{11})^i{}_j &=&  \delta^{i}{}_j\left(\lambda_1\phi^{\dagger}\phi
+\lambda_3\xi^{\dagger}\xi
+\lambda_6 (\phi^{\dagger}\xi +\xi^{\dagger}\phi)\right)\nn
&+& \lambda_1\phi^{\dagger i}\phi_{j}+ \lambda_4\xi^{\dagger i}\xi_{j}
+\lambda_6 (\phi^{\dagger i}\xi_{j} +\xi^{\dagger i}\phi_{j})\
\eea

Since we are neglecting gauge and Yukawa terms we have no
contribution to $\gamma$,  we have from~\eqn{eq:rgone}\ that
\bea
\frac{1}{2}\,\STr M^4 &=& \frac{1}{2}\,\left(\Tr A^2 + 2\Tr BC
+ \Tr D^2 \right) \nn &=&
\frac{\beta_{\lambda_1}}{2}\,\left(\phi^\dagger\phi\right)^2\,+\,
\frac{\beta_{\lambda_2}}{2}\,\left(\xi^\dagger\xi\right)^2\,+\,
\beta_{\lambda_3}\,\left(\phi^\dagger\phi\right)\,\left(\xi^\dagger\xi\right)\,+
\, \beta_{\lambda_4}\,\left(\phi^\dagger\xi\right)\,
\left(\xi^\dagger\phi\right)\, \nn
 &+& \left\{\frac{\beta_{\lambda_5}}{2}\,\left(\phi^\dagger\xi\right)^2\,+\,
 \left[\beta_{\lambda_6}\,\left(\phi^\dagger\phi\right)\,+\,
 \beta_{\lambda_7}\,\left(\xi^\dagger\xi\right) \right]\,
 \left(\phi^\dagger\xi \,+\,\mbox{h.c.}\right)\right\} \;\;\; .
\label{eq:rpot}
\eea
It is straightforward algebra to obtain from \eqn{eq:rpot}\ that
\bea \beta_{\lambda_1} &=& 12\lambda_1^2 + 4\lambda_3^2 +
4\lambda_3\lambda_4+2\lambda_4^2+2\lambda_5^2+24\lambda_6^2\nn
\beta_{\lambda_2} &=& 12\lambda_2^2 + 4\lambda_3^2 +
4\lambda_3\lambda_4 +2\lambda_4^2+2\lambda_5^2+24\lambda_7^2\nn
\beta_{\lambda_3} &=& (\lambda_1+\lambda_2)(6\lambda_3+2\lambda_4)
+4\lambda_3^2+2\lambda_4^2 +2\lambda_5^2+
4\lambda_6^2+16\lambda_6\lambda_7+4\lambda_7^2\nn \beta_{\lambda_4}
&=& 2(\lambda_1+\lambda_2)\lambda_4 +8\lambda_3\lambda_4
+4\lambda_4^2
+8\lambda_5^2+10\lambda_6^2+4\lambda_6\lambda_7+10\lambda_7^2\nn
\beta_{\lambda_5} &=& 2(\lambda_1+\lambda_2)\lambda_5
+8\lambda_3\lambda_5
+12\lambda_4\lambda_5+10\lambda_6^2+4\lambda_6\lambda_7+10\lambda_7^2\nn
\beta_{\lambda_6} &=& 12\lambda_1\lambda_6
+6\lambda_3(\lambda_6+\lambda_7)
+8\lambda_4\lambda_6+4\lambda_4\lambda_7+10\lambda_5\lambda_6+2\lambda_5\lambda_7
\nn \beta_{\lambda_7} &=& 12\lambda_2\lambda_7
+6\lambda_3(\lambda_6+\lambda_7)
+4\lambda_4\lambda_6+8\lambda_4\lambda_7
+2\lambda_5\lambda_6+10\lambda_5\lambda_7. \label{betalams} \eea
We have verified the above results by a
standard Feynman diagram calculation. Moreover they are in full agreement
with the results presented in Appendix~A of \reference{Haber:1993an},
except for a typo in the result for $\beta_{\lambda_2}$ there;
the contribution $12\lambda_6^2$ there should read $12\lambda_7^2$.
(Note that there is an overall difference of a factor of 2 between
the definitions of all the $\beta$-functions).

The contributions to $\STr M^4$ of $O(h^4)$ and $O(g^4,g^2 g^{\prime
2},g^{\prime 4})$ are easily calculated. The $O(g^4,g^2 g^{\prime
2},g^{\prime 4})$ terms come from the  gauge boson mass matrix:
\be
M_V^2 = \left(\begin{array}{cc}
\frac{1}{4}g^2\phi^{\dag}\left\{\tau^a,\tau^b\right\}\phi + (\phi
\to \xi)
&\frac{1}{2}g g^{\prime}\phi^{\dag}\tau^a\phi + (\phi \to \xi) \\
\frac{1}{2}g g^{\prime}\phi^{\dag}\tau^a\phi + (\phi \to \xi)
&\frac{1}{2}g^{\prime 2}\phi^{\dag}\phi + (\phi \to \xi)
\end{array}\right) \;\;\; .
\ee
Then using the identities
\bea
\left\{\tau^a,\tau^b\right\} &=& 2\delta^{ab}\nn
(\tau^a)^i_j (\tau^b)^k_l &=& 2 \delta^i_l\delta^k_j - \delta^i_j \delta^k_l
\eea
one easily shows that
\bea S\Tr M_V^4 &=& 3(\frac{3}{4}g^4+\frac{1}{4}g^{\prime 4}) \left(
(\phi^{\dag}\phi)^2 +
(\xi^{\dag}\xi)^2+2\phi^{\dag}\phi\xi^{\dag}\xi\right)\nn &+&
\frac{3}{2}g^2g^{\prime 2} \left( (\phi^{\dag}\phi)^2 +
(\xi^{\dag}\xi)^2 +4\phi^{\dag}\xi\xi^{\dag}\phi -2
\phi^{\dag}\phi\xi^{\dag}\xi\right) \;\;\; ,
\label{eq:vectorm}
\eea
where the $\STr$ has contributed a spin factor of $3$.

The $O(h_t^4)$ contributions come from the top mass matrix:
\be M_t^2 = h_t^2 \phi^{\dagger}\phi \ee so that \be \STr M_t^4 =
-12 h_t^4 (\phi^{\dagger}\phi)^2
\label{eq:fermionm}
\ee
where the
$-12$ consists of a colour factor of $3$ and a spin factor of $-4$.

The remaining contributions of $O(\lambda_i h_t^2, \lambda_i
g^2,\lambda_i g^{\prime 2})$ come from the anomalous dimension term
in~\eqn{eq:rgone}, the anomalous dimensions in the Landau gauge
being
\bea \gamma_{\phi} &=& 3h_t^2
-\frac{9}{4}g^2-\frac{3}{4}g^{\prime 2}\nn \gamma_{\xi} &=&
-\frac{9}{4}g^2-\frac{3}{4}g^{\prime 2}
\label{eq:gams}
\eea
Armed with these results one easily shows from
\eqnss{eq:vectorm}{eq:fermionm}{eq:gams}\ that \eqn{betalams}\
receives the following additional contributions:
\bea
\beta_{\lambda_1} &\to& \beta_{\lambda_1} + \frac{3}{4}(3g^4 +
g^{\prime 4} +2g^2 g^{\prime 2}) - 3\lambda_1 (3g^2 +g^{\prime
2}-4h_t^2)-12h_t^4  \nn \beta_{\lambda_2} &\to& \beta_{\lambda_2} +
\frac{3}{4}(3g^4 + g^{\prime 4} +2g^2 g^{\prime 2}) -3\lambda_2
(3g^2 +g^{\prime 2})  \nn \beta_{\lambda_3} &\to& \beta_{\lambda_3}
+ \frac{3}{4}(3g^4 + g^{\prime 4} -2g^2 g^{\prime 2}) - 3\lambda_3
(3g^2 +g^{\prime 2}-2h_t^2)  \nn \beta_{\lambda_4} &\to&
\beta_{\lambda_4} +3g^2 g^{\prime 2} - 3\lambda_4 (3g^2 +g^{\prime
2}-2h_t^2) \nn \beta_{\lambda_5} &\to& \beta_{\lambda_5} -
3\lambda_5 (3g^2 +g^{\prime 2}-2h_t^2)  \nn \beta_{\lambda_6} &\to&
\beta_{\lambda_6} - 3\lambda_6 (3g^2 +g^{\prime 2}-3h_t^2) \nn
\beta_{\lambda_7} &\to& \beta_{\lambda_7} - 3\lambda_7 (3g^2
+g^{\prime 2}-h_t^2).
\label{eq:fullbetas}
\eea
We also require the
one-loop $\beta$-functions for $g, g^{\prime}, g_3$ and $h_t$, which
are given by
\bea \beta_{g^{\prime}} &=& 7 g^{\prime 3}\nn \beta_{g}
&=& -3 g^3\nn \beta_{g_3} &=& -7 g_3^3\nn \beta_{h_t} &=&
h_t\left[\frac{9}{2}h_t^2 - \frac{17}{12}g^{\prime 2} -
\frac{9}{4}g^2 -8g_3^2\right].
\eea

\section{Renormalization group invariance and the exceptional region of parameter space}
\label{sec:aph}

As mentioned in section~\ref{sec:2hdm}, one can impose various
symmetries on the 2HDM in order to obtain interesting physical
consequences. It was recently proved~\cite{iva2} that the 2HDM potential
can in fact only possess six distinct symmetries
(including both discrete and continuous symmetries). This statement,
however, hides a problem: since physical predictions cannot depend
on the basis chosen for the Higgs doublets, the form of the potential
- its specific combination of parameters and respective values - is
not uniquely determined. For instance, we discussed the 2HDM
potential with a $Z_2$ symmetry; nevertheless, a potential with a
permutation symmetry between $\Phi_1$ and $\Phi_2$ - that is, a 2HDM
potential invariant under the transformation $\Phi_1\,\leftrightarrow
\,\Phi_2$ - has exactly the same physical predictions. The reason is
that both potentials, and both symmetries, are related by a basis
change on the scalar doublets (see, for instance, \cite{3hdm}).

The question then arises, how does one know whether a potential has
a given symmetry, since that symmetry can appear in an endless
number of ways in different bases? The answer is, one builds
basis-invariant quantities, the values of which reveal which,
if any, symmetries the potential has. There has been considerable
attention to developing techniques to build basis
invariants~\cite{cpi}, first as a means of detecting CP violation
and more recently to detect other types of symmetries~\cite{gh,cp},
such as $U(1)$ or $Z_2$. In fact, the authors of Ref.~\cite{gh}
built a set of basis-invariant quantities which can distinguish
between these two symmetries. However, that method failed to
identify the presence of a continuous symmetry for a given
combination of parameters, the so-called exceptional region of
parameter space (ERPS): namely,
\begin{eqnarray}
m_{22}^2 = m_{11}^2, & \hspace{4ex} & m_{12}^2=0,
\nonumber\\
\lambda_2 = \lambda_1, & \hspace{4ex} & \lambda_7 = -
\lambda_6\;\;\;.
\label{ERPS}
\end{eqnarray}
Of particular relevance is the fact that the ERPS is not a
zero-measure set of parameters, but is itself instead attained
through the imposition, on the potential, of a given
symmetry~\cite{gh}. In fact, several possible symmetries lead into
the ERPS: either combinations of discrete symmetries or generalised
CP symmetries (generalised in the sense that they do not satisfy
$\hbox{CP}^2\,=\,1$). In fact, in~\cite{cp} two generalised CP symmetries
were identified, which lead into the ERPS: one was a discrete CP
symmetry (dubbed CP2 in that reference), the other a continuous one
(dubbed CP3). The problem of identifying the presence of a
continuous symmetry was solved in Ref.~\cite{cp}, with a new basis
invariant quantity $D$, which is written, in the ERPS, in a basis
where all potential parameters are real, as
\begin{equation}
D = - \tfrac{1}{27} \left[ \lambda_5 (\lambda_1 - \lambda_3 -
\lambda_4 + \lambda_ 5)
       - 2 \lambda_6^2 \right]^2
\left[ (\lambda_1 - \lambda_3 - \lambda_4 - \lambda_ 5)^2
       + 16 \lambda_6^2 \right].
       \label{eq:D}
\end{equation}
As shown in~\cite{cp}, once in the ERPS the basis-invariant
condition $D\,=\,0$ indicates the presence of a continuous $U(1)$
symmetry. If $\lambda_6 = 0$, then $D\,=\,0$ gives the following
conditions:
\begin{equation}
\lambda_5 = 0, \hspace{4ex} \lambda_5 = \pm (\lambda_1 - \lambda_3 -
\lambda_4). \label{L6EQ0}
\end{equation}
If $\lambda_6 \neq 0$, then $D\,=\,0$ corresponds to
\begin{equation}
2 \lambda_6^2 = \lambda_5 (\lambda_1 - \lambda_3 - \lambda_4 +
\lambda_ 5). \label{L6NEQ0}
\end{equation}
In Ref.~\cite{cp} the RG invariance of the $D\,=\,0$ condition
itself was demonstrated, through a calculation which managed to
avoid using the explicit form of the $\lambda_i$ $\beta$-functions.
Using the explicit form of the $\beta$-functions of
Appendix~\ref{sec:ap}, it is simple to demonstrate that:
\begin{itemize}
\item The conditions on the quartic couplings that define the ERPS
are RG invariant. Explicitly, we see that:
\begin{itemize}
\item If $\lambda_1\,=\,\lambda_2$ and $\lambda_6\,=\,-\lambda_7$,
then we will have $\beta_{\lambda_1}\,=\,\beta_{\lambda_2}$ and
$\beta_{\lambda_6}\,=\,-\beta_{\lambda_7}$.
\end{itemize}
\item Each of the conditions in Eqs.~\eqref{L6EQ0}
and~\eqref{L6NEQ0} are RG invariant, if we are in the ERPS (that is,
with $\lambda_2 = \lambda_1$ and $\lambda_7 = -\lambda_6$). Namely,
\begin{itemize}
\item If $\lambda_5 \,= \lambda_6 \,=\, 0$, then
$\beta_{\lambda_5} \,=\, \beta_{\lambda_6} \,=\, 0$;
\item If $\lambda_6 \,=\, 0$ and $\lambda_5 = \pm (\lambda_1 -
\lambda_3 - \lambda_4)$, then we have $\beta_{\lambda_5} = \pm
\left(\beta_{\lambda_1} - \beta_{\lambda_3} -
\beta_{\lambda_4}\right)$ and $\beta_{\lambda_6} = 0$;
\item Finally, and much in the same
manner, if $\xi\,=\,2 \lambda_6^2 \,-\, \lambda_5 (\lambda_1 -
\lambda_3 - \lambda_4 + \lambda_ 5)\,=\,0$ then we also have
$\beta_\xi \,=\,0$.
\end{itemize}
\end{itemize}
However, there is a detail which must be mentioned: the above is
true if one sets the Yukawa coupling $h_t$ equal to zero - that is,
if the theory does not couple to fermions (the {\it gauge coupling\/} contributions
from \eqn{eq:fullbetas}\ may be included, however). In fact, it is extremely
difficult, if not impossible, to couple the two doublets to fermions
in a phenomenological acceptable manner, if the symmetries which
lead to the ERPS are in place. In Ref.~\cite{mani}, for instance,
the authors managed to couple the fermion sector and the scalar one
in the presence of a CP2 symmetry, but those couplings implied
masslessness for the two first generations.

\end{document}